\documentclass[12pt]{article}
\usepackage{epsfig}

\title{Adhesion of
membranes with competing specific and generic interactions}
\author{
Thomas R.~Weikl$^{1,}$\thanks{Present address: Department of Pharmaceutical
Chemistry, University of California, San Francisco, California 94143--1204,
USA} ,
David Andelman$^2$, Shigeyuki Komura$^3$, \\and Reinhard Lipowsky$^1$\\[0.2cm]
\hspace*{-1cm}  \small $^1$ MPI f\"ur Kolloid-- und Grenzfl\"achenforschung,
         14424 Potsdam, Germany  \\
  \small $^2$ School of Physics and Astronomy, Raymond and Beverly
  Sackler Faculty of Exact Science\\
  \small Tel Aviv University, Tel Aviv 69978, Israel \\
  \small $^3$  Department of Chemistry,
         Tokyo Metropolitan University, Tokyo 192-0397, Japan}

\begin{document}
\date{April 14, 2002}

\maketitle

\begin{abstract}

Biomimetic membranes in contact with a planar substrate or a second
membrane are studied theoretically. The membranes contain specific
adhesion molecules (stickers) which are attracted by the second surface.
In the absence of stickers, the trans--interaction between the membrane and
the second surface is assumed to be repulsive at short separations.
It is shown that the interplay of specific attractive
and generic repulsive interactions can lead to the formation of a
potential barrier. This barrier induces a line tension
between bound and unbound membrane segments which results in
lateral phase separation during adhesion.
The mechanism for adhesion--induced phase separation is rather general,
as is demonstrated by considering two distinct cases involving:
(i) stickers with a linear attractive potential,
 and (ii) stickers with a short--ranged square--well potential.
In both cases, membrane fluctuations reduce the potential barrier
and, therefore, decrease the tendency of phase separation.

\end{abstract}

\vspace*{1cm}


\section{Introduction}

The adhesion of biomimetic membranes and vesicles is governed by
various generic and specific interactions \cite{liposack}. The
simplest systems are provided by lipid bilayers which contain
only one or a few lipid components and which have a laterally
uniform composition. The generic interactions between one
such membrane and another surface (or between two such
membranes) can be of enthalpic or entropic origin.

The enthalpic contribution  arises  from the intermolecular forces
such as hydration,  van der Waals, and electrostatic
forces. This contribution, called the direct membrane
interaction,  can be measured if the  membrane is  essentially
flat or planar. Experimentally, a flat state can be  prepared  by
immobilizing the membrane  on a  solid substrate. Theoretically,
this state corresponds to the limit of a  large membrane rigidity.

In aqueous solution,   bilayer membranes are often quite flexible
and then exhibit  thermally--excited  undulations which  act to
renormalize their direct interaction. If the direct
interaction is purely repulsive, the undulations lead to a free
energy contribution  which can be interpreted as an entropic or
fluctuation--induced force as  proposed in
Ref.~\cite{helf78}. If the direct interaction  contains an
attractive potential well,  the renormalized interaction  leads
to an unbinding transition  as   predicted in
Ref.~\cite{lipo33}.

Biomembranes contain a large number of different lipids and
anchored macromolecules. The attractive forces between two
membranes are usually mediated by  receptors or adhesion
molecules which are anchored in the membranes
\cite{bellEtAl,braunEtAl,brui94}. These specific interactions
govern the highly selective binding of cells which is essential
for  many biological processes such as  immune response or
tissue development \cite{albe94}. From a theoretical point of
view, the adhesive behavior of these rather complex biomembranes
can be modelled, to a certain extent, by two--component membranes
containing  `generic' lipids  and anchored   stickers
\cite{lipo130,lipo176}.

The interplay of generic and specific interactions has also been
investigated  experimentally.  Adhesion--induced  lateral phase
separation into domains with small and large membrane separations  has
been found to occur in several  biomimetic systems. The formation of
blisters has been observed in membranes containing cationic lipids in
contact with a negatively charged surface \cite{nard97}, and between
membranes containing both negatively and positively charged lipids
\cite{wong01}.  The coexistence of tight and weak adhesion states has
been found for membranes with biotinylated lipids  bound to another
biotinylated surface via streptavidin \cite{albe97},  membranes with
homophilic csA--receptors from the slime mold {\it Dictyostelium
discoideum} \cite{klob99}, and   membranes containing specific ligands
of integrin molecules adsorbed on a substrate
\cite{gutt01}. Attractive  membrane-mediated interactions between
bound csA--receptors of adhering  vesicles have been  inferred from
membrane tension jumps induced by the  micropipet aspiration technique
\cite{maie01}. In addition to the receptors,  the membranes studied in
\cite{albe97,klob99,gutt01,maie01} also contain  repulsive
lipopolymers to prevent non--specific adhesion.  These  observations
indicate the existence of possible physical mechanisms for  the
aggregation of receptors in biological membranes which has been found
during cell adhesion and results in the formation of focal contacts
\cite{albe94}.

In this article, we present a detailed study of one possible
mechanism for adhesion--induced phase separation concerning
membranes with generic repulsive and specific attractive
interactions. For these membranes, tracing over the sticker
degrees of freedom in the partition function leads to an
effective trans--membrane interaction which exhibits a potential
barrier. Such trans--membrane interactions have been previously
studied in Refs.~\cite{lipo118,lipo124}. The phase separation
results from the line tension between bound and unbound membrane
regions due to this potential barrier. Membrane fluctuations
reduce the barrier and, therefore, decrease the tendency for
phase separation. The mechanism is thus clearly distinct from
entropic, fluctuation--induced mechanisms for adhesion--induced
phase separation as discussed in
Refs.~\cite{lipo176,weiklNetzLipowsky}. Similar mechanisms for
phase separation due to an effective potential barrier have also
been studied recently in
Refs.~\cite{lipo176,komuraAndelman,brui00}.

The mechanism studied here is rather general as will be demonstrated
by considering two different cases.
In the first case, we assume that the generic trans--membrane interaction
is repulsive for short separations and attractive
for large separations. The generic trans--interaction is then approximated
by a harmonic potential centered at the potential minimum at $l = l_o$. The
specific trans--interaction is expanded up to linear order in $l-l_o$,
 and can be thought to arise from
restoring forces of extensible sticker molecules which are irreversibly
bound to the membrane and the substrate. In the absence of membrane
shape fluctuations, the lateral phase behavior can be determined
exactly. Using Monte Carlo simulations, we furthermore show
that the membrane fluctuations reduce the potential
barrier and the tendency for lateral phase separation.

In the second case, we consider a   generic
trans--membrane repulsion which is modeled as a square--barrier
potential. In addition, the trans--interaction of
the stickers is modeled as a square--well potential
\cite{lipo130,lipo176,weiklNetzLipowsky}. The stickers are bound
for small separations from the substrate, and unbound otherwise.
The square--well potential is a simple model for
short--ranged lock--and--key interactions of ligands and receptors.
After partial summation over the sticker degrees of freedom, we find again
an effective potential barrier if the generic repulsion
is longer ranged than the specific attraction of the stickers.
As in the first case, this barrier leads to
lateral phase separation, and is effectively reduced by membrane
fluctuations.

\section{The model}

\begin{figure}
\begin{center}
\epsfig{figure=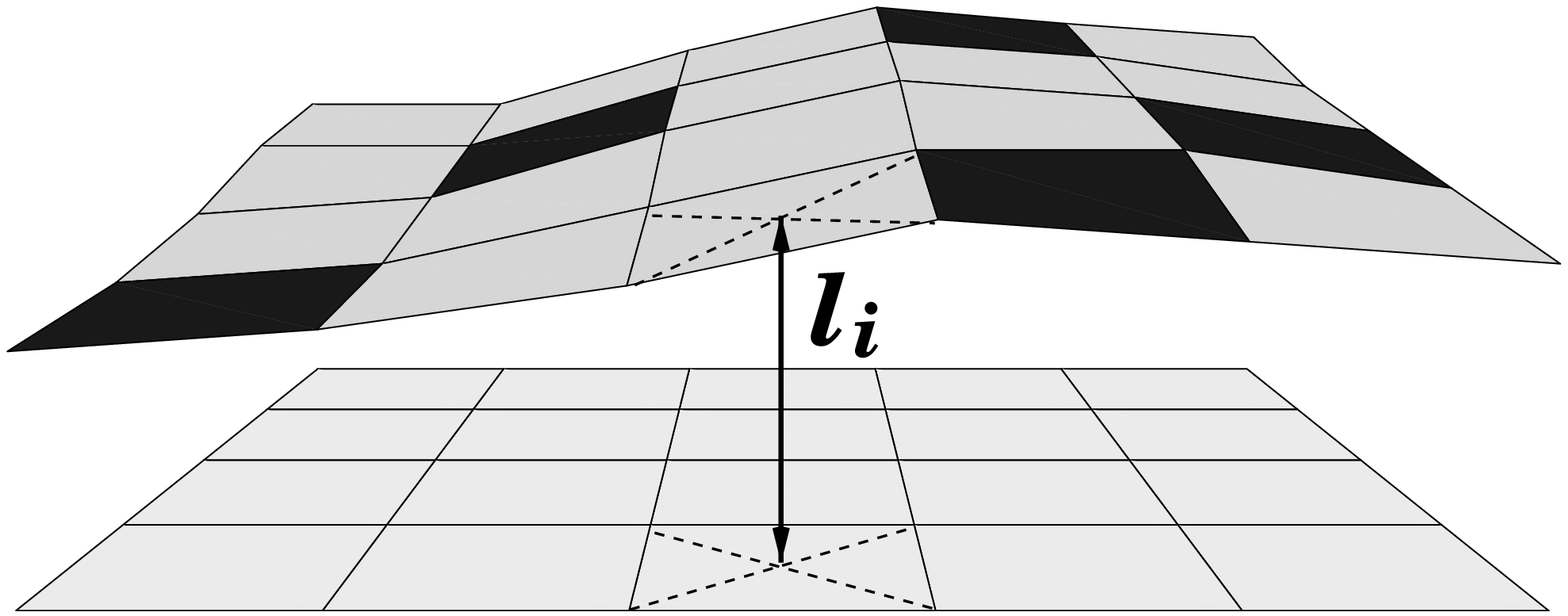, width=1\linewidth}
\end{center}
\vspace{1cm} \caption{A membrane segment consisting of $5 \times
4$ membrane patches in contact with another planar surface. The
patches are labeled by the lattice sites $i$. The local
separation of the membrane from the reference plane  is denoted
by  $l_i$. The composition of the membrane is described by
occupation numbers $n_i = 0$ and $n_i = 1$ corresponding to grey
patches with no sticker  and black patches with one sticker,
respectively. } \label{modelFig}
\end{figure}

A systematic description of a biomimetic membrane with stickers in
contact with a substrate or a second membrane has to include a
field $l(x,y)$ for the local separation of the membrane(s) and a
concentration field $n(x,y)$ of the  stickers above a position
$(x,y)$ on a reference plane. In the following, we apply a
theoretical framework which has been introduced in
Ref.~\cite{lipo130} and extended in
Refs.~\cite{lipo176,weiklNetzLipowsky}. Within this framework, the
membrane is divided into small patches with a linear size $a$
which corresponds to the smallest possible wavelength for bending
deformations. According to computer simulations for molecular
membranes, this size is about 6 nm for lipid bilayers with a
thickness of about 4 nm \cite{lipo154}. For a membrane which is
on average parallel to a planar substrate, this leads to an
effective discretization of the reference plane into a
two--dimensional square lattice with lattice parameter $a$.
 The sticker positions   are described by occupation
numbers $n_i=0$ or 1 where $n_i=1$ indicates the presence of a
sticker at lattice site $i$, and the local separation is given by
$l_i\ge 0$ \cite{lipo130}, see Fig.\ \ref{modelFig}. An
alternative continuous Ginzburg--Landau theory for the sticker
concentration field  was used in Ref.~\cite{komuraAndelman}.

In terms of these variables, the grand canonical Hamiltonian
has the general form
\begin{equation}
{\cal H}\{l,n\} = {\cal H}_{el}\{l\} + \sum_i \left[V_g(l_i) + n_i
\left(V_s(l_i)-\mu\right)\right]  \label{genham}
\end{equation}
where the elastic term
\begin{equation}
{\cal H}_{el}\{l\}=\sum_i\frac{\kappa}{2 a^2}(\Delta_d l_i)^2
\end{equation}
represents the discretized bending energy of the membrane with
bending rigidity $\kappa$, and the discretized Laplacian
$\Delta_d$ is given by
\begin{equation}
\Delta_d l_i=\Delta_d l(x,y) = l(x+a,y)+l(x-a,y)+l(x,y+a)+l(x,y-a)
-4l(x,y) \label{laplacian}
\end{equation}
The term $(\Delta_d l_i )^2$ corresponds to the leading order
expression for the mean curvature squared of a membrane with
vanishing spontaneous curvature \cite{canham,helf73}. The
second term of the Hamiltonian (\ref{genham}) contains (i) the
generic interaction potential $V_g(l)$ between the membrane
and substrate and (ii) the specific adhesion potential $V_s(l)$ of
the stickers which only contributes at lattice sites with
$n_i=1$, {\it i.e.}~at lattice sites where stickers are present.
The relative chemical potential of the  stickers  is denoted by
$\mu$. The same description holds for a multicomponent membrane
in contact with a second, homogeneous membrane. In the latter case,
the effective bending rigidity $\kappa$ is  given by $\kappa_1
\kappa_2/(\kappa_1 + \kappa_2)$ where $\kappa_1$ and $\kappa_2$
denote the bending rigidities of the two membranes \cite{rlEPL88}.

Note that $V_g(l)$ and  $V_s(l)$ are the interaction energies
{\em per membrane patch} where each patch has   area $a^2$.
  Thus, the  interaction energies per unit area are  given by
$V_g(l)/a^2$ and  $V_s(l)/a^2$, respectively.  This differs from
the convention in Ref.~\cite{lipo130} where the interaction
potentials were defined as  energies per unit area.

In the following, we neglect direct interactions between
pairs of stickers which can be described by quadratic terms in the
concentration field $n$ \cite{lipo176,weiklNetzLipowsky}.
The Hamiltonian (\ref{genham}) is then   linear in $n$, and the sticker
degrees of freedom in the partition function
\begin{equation}
{\cal Z}=  \bigg[\prod_i\int_{0}^{\infty} {\rm d}l_i\bigg]
\bigg[\prod_i\sum_{n_i=0,1}\bigg]
  \exp\left[-\frac{{\cal{H}}\{l,n\}}{T}\right]
\end{equation}
can be partially summed or traced over exactly, leading to
\begin{equation}
{\cal Z}=   \bigg[\prod_i\int_{0}^{\infty} {\rm d}l_i\bigg]
  \exp\left[-\frac{{\cal H}_{el}\{l\}+\sum_i V_{ef}(l_i)}{T}\right]
\end{equation}
with the effective potential
\begin{equation}
 V_{ef}(l) =  V_g(l)
         -T\ln\left(1+\exp\left[ \frac{\mu - V_s(l)}{T}\right] \right)
\label{Vef}
\end{equation}
where $T$ denotes the temperature in energy units (i.e., the
Boltzmann constant $k_B$ is absorbed into the symbol $T$).
The partial summation over the sticker degrees of freedom $\{n\}$
thus leads to  an equivalent problem of a laterally {\em
homogeneous} membrane with the effective potential (\ref{Vef}).

\section{Linear sticker potential}

In this section, we consider a generic potential $V_g$ between the
membrane and the substrate which has a relatively deep minimum at a certain
separation $l_o$ from the substrate. Such a potential  arises, {\em e.g.},  for
electrically neutral surfaces interacting via strong van der Waals forces.
Using a Taylor expansion around the minimum,  we approximate
this generic potential by the harmonic potential
\begin{equation}
V_g(l) = \frac{v_2}{2a^2} (l - l_o)^2
\label{harmpot}
\end{equation}
where $v_2 = a^2 (d^2 V_g/dl^2)|_{l_o}$.

In addition to this generic potential, we   consider extensible sticker
molecules which are irreversibly bound to both the
substrate and the membrane, and which have an unstretched
extension small compared to $l_o$. We will further assume that
the corresponding  sticker potential  $V_s(l)$ has an
essentially constant
gradient for those values of $l$ for which we can use the
harmonic approximation (\ref{harmpot}) for the generic potential.
In such a situation, we may truncate the expansion of
the sticker potential in powers  of  $l-l_o$ and use
\begin{equation}
V_s(l) = V_s(l_o) + \frac{\alpha (l - l_o)}{a}
\label{linearpot}
\end{equation}
with  $\alpha \equiv a \partial V_s(l)/ \partial l |_{l_o} >  0$.

In order to simplify the notation and to reduce the number of parameters,
it is convenient to introduce the rescaled variables
\begin{equation}
h \equiv \sqrt{\frac{v_2}{T}} \;\frac{l - l_o}{a}
\end{equation}
The Hamiltonian (\ref{genham}) with the generic potential
(\ref{harmpot}) and the specific potential (\ref{linearpot}) can
then be written as
\begin{equation}
\frac{ {\cal H}\{h,n\}}{T} =
 \sum_i\left[\frac{\kappa}{2 v_2}(\Delta_d h_i)^2
+\frac{1}{2}h_i^2 + n_i \left(\tilde{\alpha}h_i -\tilde{\mu}
   \right)\right]  \label{linearham}
\end{equation}
in terms of the discrete lattice variables $h_i$, $n_i$,
and the dimensionless parameters
\begin{equation}
\tilde{\alpha} = \frac{\alpha}{\sqrt{v_2T}} \hspace{0.5cm}
\mbox{and} \hspace{0.5cm} \tilde{\mu}=\frac{\mu - V_s(l_o)}{T}
\quad ,
\label{AlphaReduced}
\end{equation}
and the effective potential (\ref{Vef})  has the form
\begin{equation}
\frac{V_{ef}(h)}{T} =   \frac{1}{2}h^2
  -\ln\left(1+e^{\tilde{\mu} -\tilde{\alpha}h}\right)
\label{VefII}
\end{equation}
Direct inspection of these equations shows that the system
considered here depends on
 three dimensionless parameters: (i)  the reduced coupling
constant $\tilde{\alpha}$ of the specific potential, (ii) the
reduced (and shifted) chemical potential  $\tilde{\mu}$,  and (iii) the ratio
$\kappa/v_2$ of the bending rigidity $\kappa$ and the strength
$v_2$ of the generic harmonic potential (\ref{harmpot}).

\subsection{Limit of rigid membranes}

For large values of  the ratio $\kappa/v_2$, thermally excited
shape fluctuations of the membrane can be neglected. The free
energy $F = - (T/A)\ln {\cal Z}$  per area $A$ is then given by
$V_{ef}/a^2$. The phase behavior is determined by the
minimization of the effective potential (\ref{VefII}):
\begin{equation}
\frac{\partial V_{ef}}{\partial h}=0
\end{equation}
First-order phase transitions are found when
different minima of $V_{ef}$ coexist.

For the effective potential (\ref{VefII}),
the  discussion of the phase behavior
is simplified if this potential is expressed in
terms of the shifted  separation field
\begin{equation}
z\equiv h +\tilde{\alpha}/2
\quad .
\label{shift}
\end{equation}
For the special line in the $(\tilde{\mu}, \tilde{\alpha})$
parameter space as given by
\begin{equation}
 \tilde{\mu} = \tilde{\mu}_*\equiv -\tilde{\alpha}^2/2
\quad ,
\label{mustar}
\end{equation}
the effective potential (\ref{VefII}) has the  form
\begin{equation}
\frac{V_{ef}(z)}{T}\bigg|_{\tilde{\mu}= \tilde{\mu}_*} =
\frac{z^2}{2}+\frac{\tilde{\alpha}^2}{8}
-\ln\left[2\cosh(\tilde{\alpha}z/2)\right]
     \label{Vefstar}
\end{equation}
which is symmetric under the inversion $z\to -z$.

\begin{figure}
\hspace{2cm} \epsfig{figure=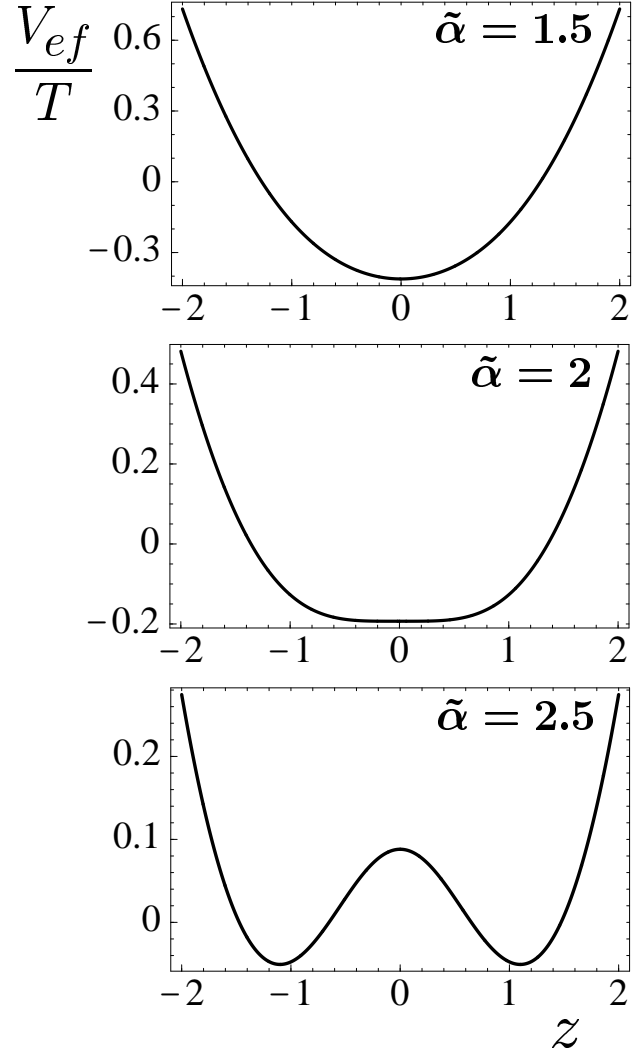, width=0.6\linewidth}
\caption{ The effective potential $V_{ef}$  as a function of the
shifted separation variable $z$ for three values of the coupling
$\tilde{\alpha}$. The analytical expression for $V_{ef}$ is given
in (\ref{Vefstar}). For small  and large values of
$\tilde{\alpha}$, $V_{ef}$ exhibits a single minimum and   two
degenerate minima, respectively, as shown in the top and bottom
parts. At   $\tilde{\alpha} =
\tilde{\alpha}_c=2$, the potential has the shape shown in the
middle  and  undergoes a continuous bifurcation. } \label{VefFig}
\end{figure}

As one varies the parameter $\tilde{\alpha}$ while keeping
$\tilde{\mu}=\tilde{\mu}_*(\tilde{\alpha})$, the effective
potential given by (\ref{Vefstar}) undergoes a continuous
bifurcation at the critical value $\tilde{\alpha}
=\tilde{\alpha}_c = 2$, see Fig.~\ref{VefFig}.  For
$\tilde{\alpha}<\tilde{\alpha}_c$ and
$\tilde{\alpha}>\tilde{\alpha}_c$, this potential has a  single
minimum at $z=0$ and two degenerate minima at finite values of
$z$, respectively.
The critical value $\tilde{\alpha}_c = 2$ of the bifurcation
point can be directly  inferred from  the  second derivative of
(\ref{Vefstar}) as given by
\begin{equation}
\frac{1}{T}\frac{d^2 V_{ef}(z)}{d z^2}
 \bigg|_{\tilde{\mu}=\tilde{\mu_*}}=
1-\frac{\tilde{\alpha}^2}{4\cosh^2(\tilde{\alpha}z/2)}
\quad .
\end{equation}
For $z=0$,
this   expression is equal to  $1-\tilde{\alpha}^2/4$
which  vanishes for $\tilde{\alpha} = \tilde{\alpha}_c
= 2$. In the limit  of rigid membranes as
considered here, one can ignore the effect of membrane
fluctuations and the bifurcation point of the effective potential
is identical with the critical point of the system which thus
lies at $\tilde{\alpha}_c = 2$ and
$\tilde{\mu}_c=-\tilde{\alpha_c}^2/2=-2$.

Thus, for $\tilde{\mu} = \tilde{\mu}_*(\tilde{\alpha})$ and
 $\tilde{\alpha}>2$, the effective potential (\ref{VefII}) is a
symmetric double--well potential with two degenerate minima.
As soon as the chemical potential  $\tilde{\mu}$ deviates from
its  coexistence value  $\tilde{\mu} = \tilde{\mu}_*$, this
symmetry is broken and the effective potential exhibits a unique
global minimum. Therefore,
 the system undergoes a discontinuous transition as one
changes the chemical potential from $\tilde{\mu} = \tilde{\mu}_*
- \epsilon$ to $\tilde{\mu} = \tilde{\mu}_* + \epsilon$
for $\tilde{\alpha}>2$ where $\epsilon$ denotes a small chemical
potential difference.

The positions,  say $z_o$, of the extrema  of the effective
potential are determined by $dV_{ef}(z)/dz=0$. Along the  coexistence
line given by $\tilde{\mu}= \tilde{\mu}_*=-\tilde{\alpha}^2/2$,
this leads to the transcendental equation
\begin{equation}
z_o = \frac{\tilde{\alpha}}{2} \tanh\left( \frac{\tilde{\alpha}z_o}{2}
      \right)
\quad .
\label{trans}
\end{equation}
This equation  has the trivial solution $z_o
=0$ for all values of $\tilde{\alpha}$ which corresponds to a  minimum and
maximum for
$\tilde{\alpha} < \tilde{\alpha}_c=2$ and
$\tilde{\alpha}>\tilde{\alpha}_c=2$, respectively.
For $\tilde{\alpha}>\tilde{\alpha}_c=2$, equation (\ref{trans}) has two
additional solutions corresponding to the two degenerate minima
of the effective potential $V_{ef}$, see
Fig.~\ref{VefFig}.

For rigid membranes with large $\kappa/v_2$,
the sticker concentration $X\equiv\langle n_i\rangle/a^2$
  is  given by
\begin{equation}
X = -\frac{\partial F}{\partial \mu}
= -\frac{1}{a^2}\frac{\partial V_{ef}}{\partial \mu}
=\frac{1}{a^2}
\frac{e^{\tilde{\mu}-\tilde{\alpha}y_o}}
    {1+e^{\tilde{\mu}-\tilde{\alpha}y_o}}
\end{equation}
with  $y_o \equiv z_o  - \tilde{\alpha}/2$ which  denotes the
position of the minima of the effective potential (\ref{VefII}).
Along the coexistence line with $\tilde{\mu} =
\tilde{\mu}_*=-\tilde{\alpha}^2/2$, this expression simplifies and
becomes
\begin{equation}
X \bigg|_{\tilde{\mu}= \tilde{\mu}_*}= \frac{1}{a^2}
\frac{e^{-\tilde{\alpha}z_o}}{1+e^{-\tilde{\alpha}z_o}}
\quad  .
\label{Xeins}
\end{equation}

\begin{figure}
\begin{center}
\epsfig{figure=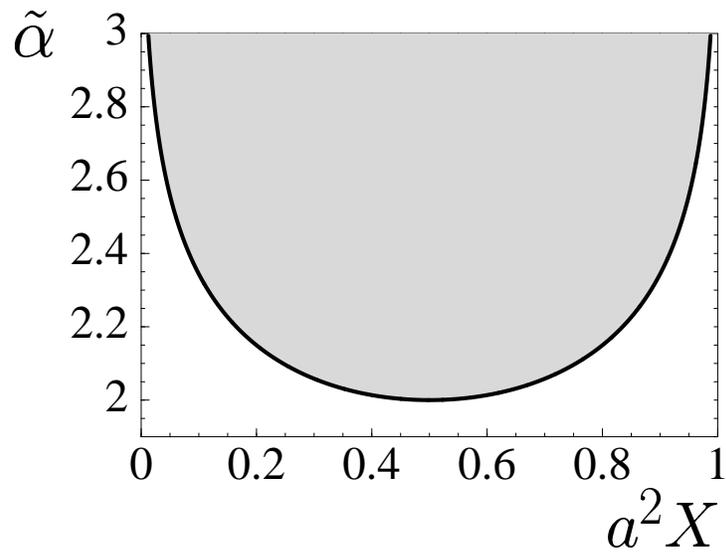, width=0.7\linewidth}
\end{center}
\caption{Phase diagram for linear stickers in the absence of membrane
fluctuations, depending on the sticker concentration $X$
and the reduced coupling constant $\tilde{\alpha}$.  Within the grey
two--phase region,    a sticker--poor phase
characterized by a relatively large membrane--surface separation coexists
with  a sticker--rich phase for which this separation is relatively small.
The critical point is located at $a^2 X_c = 1/2$ and
$\tilde{\alpha}_c = 2$.
\label{phaseDia}}
\end{figure}

Inserting the numerically determined solutions
of the transcendental equation (\ref{trans}) into (\ref{Xeins})
leads to the concentrations of the coexisting phases which determine
the phase diagram shown in Fig.~\ref{phaseDia}. Inside the shaded
two--phase region, a sticker--poor phase characterized by a
relatively large
separation $y_o$ of the membrane from the substrate coexists with a
sticker--rich phase with a  relatively small value of  $y_o$.

\subsection{Flexible membranes}

For a fluctuating membrane in a symmetric double--well potential,
first--order transitions only exist if the barrier between the two
potential wells exceeds a certain critical height
\cite{lipo118,lipo124}. For barrier heights below this critical
value, the fluctuating membrane `tunnels' through the barrier and
there is no phase transition. Therefore, the critical point of a
flexible membrane in the double--well potential (\ref{Vefstar})
will be characterized by reduced coupling constants
$\tilde{\alpha}_c=\tilde{\alpha}_c(\kappa/v_2)$ which exceed the
bifurcation value $\tilde{\alpha}_c=2$ as obtained for
rigid membranes in the limit of large $\kappa/v_2$.

\begin{figure}
\begin{center}
\epsfig{figure=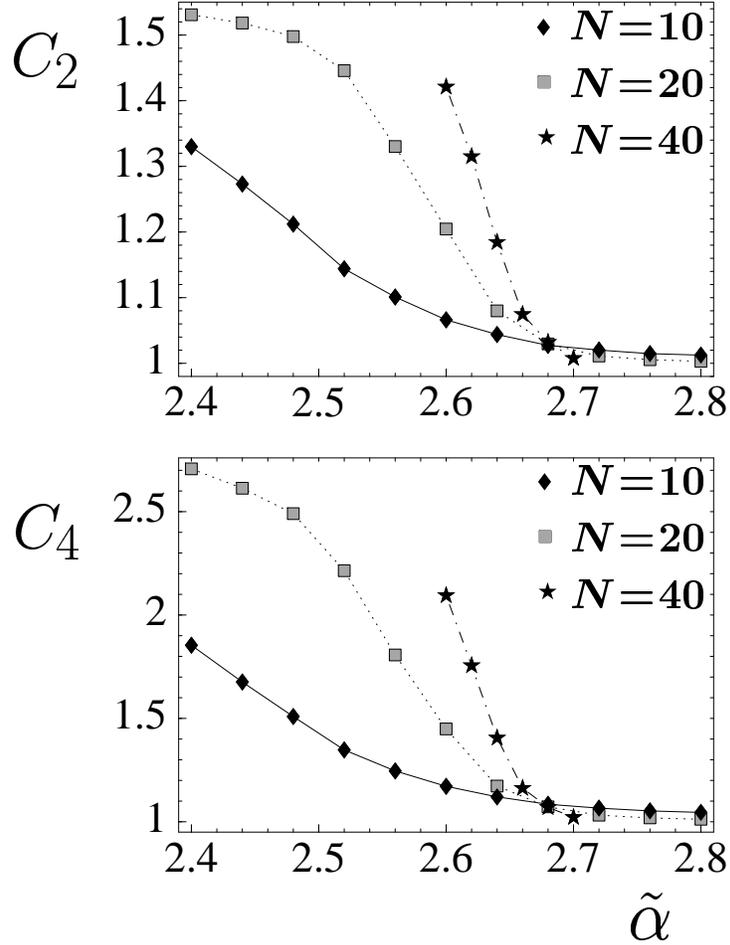, width=0.7\linewidth}
\end{center}
\caption{Monte Carlo data for the moments $C_2$ and $C_4$ defined
in (\ref{C2C4Moments}) as a function of the reduced coupling
constant $\tilde{\alpha}$. The  ratio of the bending rigidity
$\kappa$ and  of the strength $v_2$ for the generic harmonic
potential (\ref{harmpot}) has the fixed value  $\kappa/v_2 = 1$.
The membrane segments studied in the simulations consist of $N
\times N$ membrane patches.
 \label{MCcumulants}}
\end{figure}

Using Monte Carlo simulations, the critical point  can be determined,
for a fixed value of  $\kappa/v_2$,  from an evaluation of the moments
\begin{equation}
C_2 = \frac{\langle\bar{z}^2\rangle}{\langle |\bar{z}|\rangle^2}
\hspace{0.5cm} \mbox{and} \hspace{0.5cm}
C_4 = \frac{\langle\bar{z}^4\rangle}{\langle \bar{z}^2\rangle^2}
\label{C2C4Moments}
\end{equation}
where
\begin{equation}
\bar{z} = \frac{1}{N}\sum_{i=1}^N z_i
\end{equation}
is the spatially averaged order parameter, and
$\langle\cdots\rangle$ denotes averages over all membrane
configurations \cite{lipo124,binder}.
For $\tilde{\alpha}>\tilde{\alpha}_c$ and
correlation lengths $\xi$ which are much smaller than the linear
size $L$ of the finite membrane, the moments reach the values
$C_2=1$ and $C_4=1$, whereas for
$0<\tilde{\alpha}<\tilde{\alpha}_c$ and $\xi\ll L$, we have
$C_2=\pi/2\approx 1.57$ and $C_4=3$. For $L\ll\xi$ on the other
hand, the moments $C_2$ and $C_4$ vary only weakly with the
linear size $L$. The critical value $\tilde{\alpha}_c$ of the
reduced coupling constant can then be estimated from the common
intersection point  of $C_2$ and $C_4$, respectively, as  a
function of  $\tilde{\alpha}$  for several values of $L$
\cite{lipo124,binder}, see Fig.~\ref{MCcumulants}.

\begin{figure}
\begin{center}
\epsfig{figure=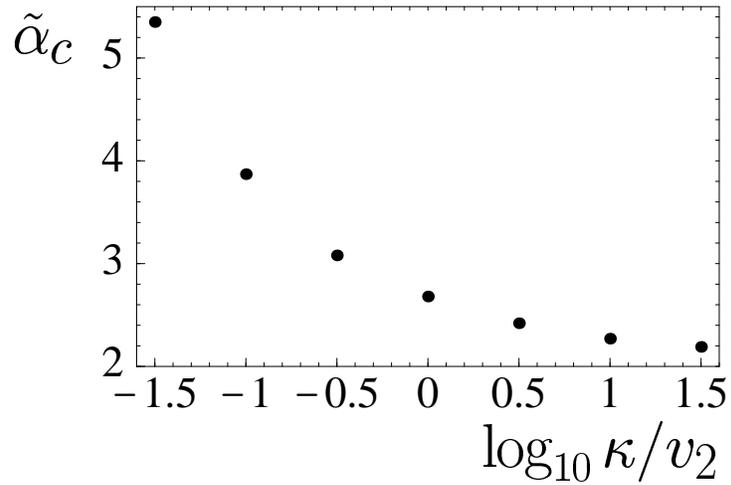, width=0.7\linewidth}
\end{center}
\caption{The critical  values   $\tilde{\alpha}_c$
of the reduced coupling constant $\tilde{\alpha}$
 as a function of the reduced membrane rigidity  $\kappa/v_2$. The
reduced  coupling
constant  $\tilde{\alpha}$ is defined in  (\ref{AlphaReduced})
and governs the
strength of the linear sticker potential as given by
(\ref{linearpot}).
For large values of $\kappa/v_2$,  one attains the limit of rigid
membranes with $\tilde{\alpha_c} \approx 2$.
The statistical errors are smaller than the size of the symbols. }
\label{MCcritpoints}
\end{figure}

In Fig.~\ref{MCcritpoints}, we display the obtained values for the
critical rescaled coupling constant $\tilde{\alpha}_c$ as a function
of the reduced rigidity $\kappa/v_2$. For large $\kappa/v_2$,
$\tilde{\alpha}_c$ approaches the limiting value
$\tilde{\alpha}_c=2$ as appropriate for rigid membranes as discussed
in the previous section. As one decreases
$\kappa/v_2$, the  membrane fluctuations become more pronounced and
act to increase the  value of
$\tilde{\alpha}_c$. This implies that the   wells of the effective
potential (\ref{Vefstar})   have a finite depth as one
reaches the critical point of the system.

Lateral phase separation occurs for coupling constants
$\tilde{\alpha}>\tilde{\alpha}_c$.
In either of the two phases, the entire membrane is located around
one of the minima in the effective potential. In the
sticker--poor phase, the membrane is found in the minimum with
larger separation $y_o$ from the substrate. This minimum is
dominated by the generic membrane potential and corresponds to a
state of weak adhesion. In the sticker--rich phase, the membrane
fluctuates around the minimum with smaller separation $y_o$
corresponding to a state of tight adhesion. In contrast, there is
only a single phase for coupling constants
$\tilde{\alpha}<\tilde{\alpha}_c$. For example, for
$2<\tilde{\alpha}<\tilde{\alpha}_c$ the two minima of the
effective potential are both populated by many different segments
of the fluctuating membrane which is able to cross the potential
barrier between the minima.

\section{Square--well sticker potential}

Let us now turn to stickers with a specific adhesion potential
\begin{eqnarray}
V_s(l) = U\theta(l_v - l) &=U & \mbox{for $0\le l \le l_v$}
                          \nonumber\\
                  &= 0 & \mbox{for $l > l_v$}   \label{squarewell}
\end{eqnarray}
where $\theta(x)$ is the Heaviside function: $\theta(x) = 0$ for
$x<0$ and $\theta(x) = 1$ for $x\ge 0$.
The parameter $U$ has the same dimension as the membrane potential
$V_s$ and represents the interaction energy per sticker.

Stickers which interact via the
square--well potential (\ref{squarewell}) can attain two
different states: A bound state with binding energy $U<0$ if the
local separation $l$ between the membrane and the substrate is
smaller than the potential range $l_v$, and an unbound state
otherwise. Because the fluctuating membrane cannot
penetrate the substrate surface, the
membrane separations $l$ are restricted to nonnegative values. In
contrast to the linear sticker potential of the previous section
with the specific potential (\ref{linearpot}), the stickers
characterized by the interaction potential (\ref{squarewell}) have a
fixed length and cannot be stretched.
Thus, the square--well potential as given by  (\ref{squarewell})
provides  a simple model for short--range interactions arising, e.g., from
specific ligand/receptor lock--and--key interactions  or from screened
electrostatic forces for charged stickers in contact with an
oppositely charged substrate.

The phase behavior of multicomponent membranes
containing  stickers which interact via the  square--well potential
as given in (\ref{squarewell}) has been  studied previously
for the case in which one can ignore   generic interactions with the
substrate,
The membrane was found to undergo  lateral phase separation even
for purely repulsive cis--interactions between the stickers if these
stickers have an increased lateral size \cite{weiklNetzLipowsky} or a
larger bending rigidity than the lipid matrix \cite{lipo176}.
The phase separation is then
driven by the shape fluctuations of the membrane.

\begin{figure}
\begin{center}
\epsfig{figure=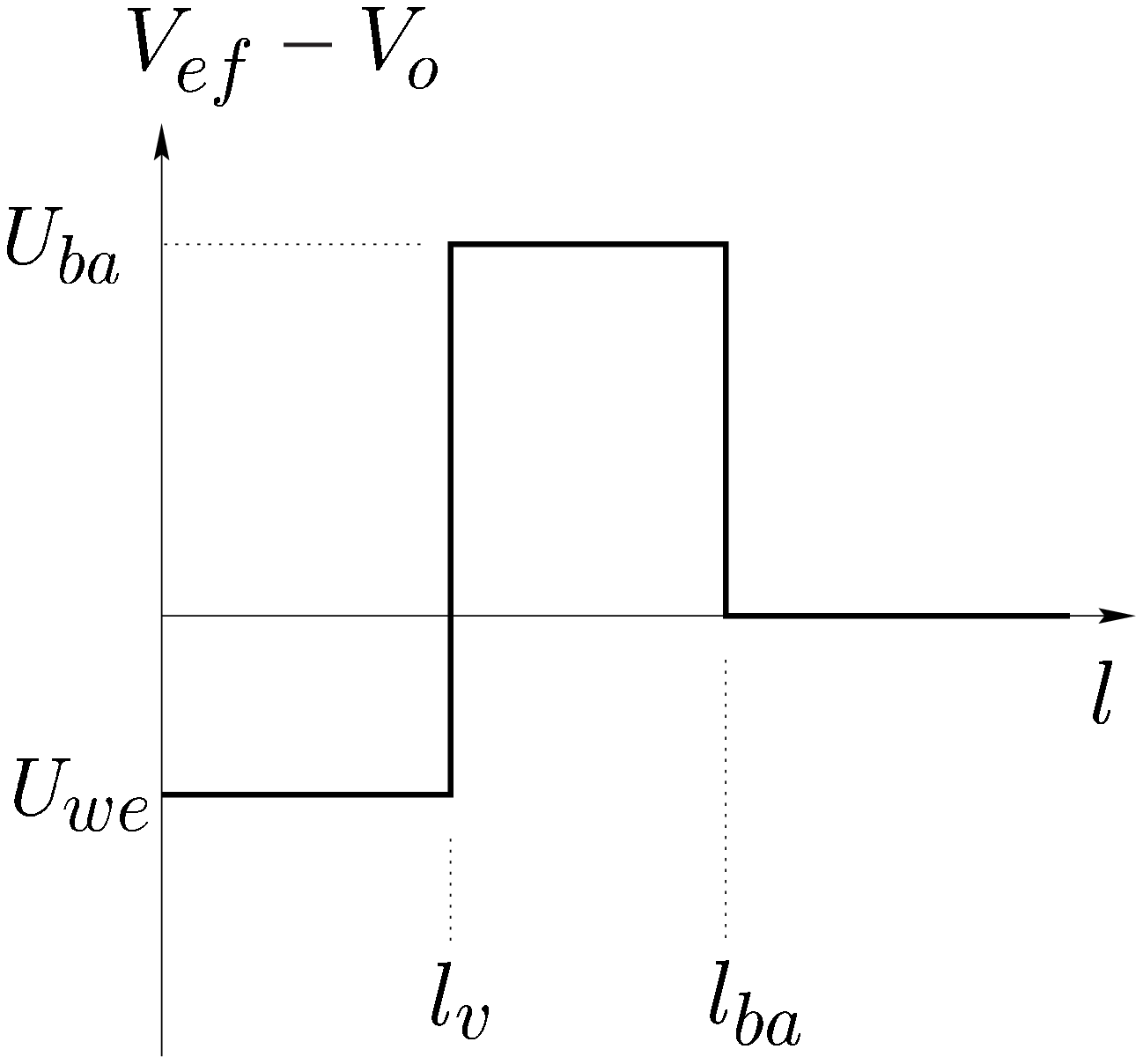, width=0.8\linewidth}
\end{center}
\caption{Schematic form of the  effective potential $V_{ef}-V_o$
defined in (\ref{potbar})
as a function of the membrane
separation $l$.  The effective potential exhibits (i) a potential barrier of
height
$U_{ba}$ which extends up to the separation $l_{ba}$, see
(\ref{genbarr}), and (ii) a potential well
which has  the range $l_v$, arising from the short--ranged sticker
 potential (\ref{squarewell}),  and the effective
depth $U_{we}$ as given by (\ref{Uco}). }
\label{potsketch}
\end{figure}

Here, we consider the interplay of the specific sticker potential
(\ref{squarewell}) with a generic repulsive trans--interaction
between the membrane and the substrate.
If the range of these generic interactions is smaller than the potential
range $l_v$ of the stickers, the bound state of the stickers is more
restricted, but the general entropic phase behavior described above will
not be affected. However, for repulsive generic interactions with a
potential range which exceeds  $l_v$, a different scenario is possible.
For simplicity, we consider here a generic potential of the form
\begin{eqnarray}
V_g(l) = U_{ba}\theta(l_{ba} - l) &=U_{ba} &
    \mbox{for $0\le l \le l_{ba}$}
                          \nonumber\\
                  &= 0 & \mbox{for $l > l_{ba}$}   \label{genbarr}
\end{eqnarray}
with a barrier height $U_{ba}>0$ and range $l_{ba}>l_v$. The
effective potential (\ref{Vef}) obtained after the summation of
the sticker degrees of freedom is shown in Fig.~\ref{potsketch}
and can be written as
\begin{eqnarray}
V_{ef}(l) -V_o=& U_{we}&  \hspace{0.2cm}  \mbox{for \  $0<l<l_v$}
 \nonumber\\
             =& U_{ba}&   \hspace{0.2cm}  \mbox{for \  $l_v<l<l_{ba}$}
  \nonumber\\
            =&  0 & \hspace{0.2cm}     \mbox{for \ $l_{ba}<l$}
   \label{potbar}
\end{eqnarray}
with
\begin{equation}
 U_{we} = U_{ba} - T\ln\frac{1+e^{(\mu - U)/T}}{1+e^{\mu /T}}
 \label{Uco}
\end{equation}
and the constant term $V_o=-T\ln\left(1+e^{\mu/T}\right)$ which
depends only on the reduced chemical potential $\mu/T$. Since the
sticker binding energy $U$ is negative, we have $U_{we}<U_{ba}$.

In the context of interacting membranes, an effective  potential
of the form (\ref{potbar}) was first studied in
Ref.~\cite{lipo118}. More recently, such a interaction
potential has also been derived for membranes which contain both
stickers and repellers, {\it i.e.}~non--adhesive molecules which
protrude from the membrane surface \cite{lipo176}. The repellers
have been modeled by a local square--barrier potential, the
stickers by the local square--well potential (\ref{squarewell}).
The generic square--barrier potential (\ref{genbarr}) thus affects
the phase behavior in a similar way as repellers with a local
square--barrier potential. As discussed previously, the membrane
unbinds at a certain critical depth
 $U_{we}^*$ of the attractive potential well which is given by
  \cite{lipo118}
\begin{equation}
|U_{we}^*| = c a^2 T^2/\kappa l_v^2 \label{Ucostar}
\end{equation}
where $c$ is a dimensionless coefficient, because the excess
free energy for a membrane confined to a potential well with
depth $U_{we}$ is  $V_{fl}(l_v) \sim T^2/\kappa l_v^2$,  and the free
energy difference between the bound and the unbound state of the
membrane reads $\Delta F=-|U_{we}| / a^2  + c T^2/\kappa l_v^2$. The
character of the unbinding transition depends on the height of  the
potential barrier which induces a line tension between bound and unbound
membrane segments. On small length scales, this line tension can be
estimated as $U_{ba}^{eff}L$ with the effective barrier height
$U_{ba}^{eff}=U_{ba}+|U_{we}|-c T^2/\kappa l_v^2$ and the effective
width $L\sim (l_{ba}-l_v)\sqrt{\kappa/T}$ of the membrane strip
crossing the barrier. \cite{lipo118} Taking also into account the line entropy
on larger length scales, the unbinding transition
is found to be discontinuous for relatively strong barriers with
\begin{equation}
 U_{ba} (l_{ba} - l_v)^2 \gg |U_{we}^*| l_v ^2 \label{barcon}
\end{equation}
and continuous for weak barriers with $U_{ba} (l_{ba} - l_v)^2
\ll |U_{we}^*|l_v ^2$.  A discontinuous
transition implies the coexistence of a bound phase with a high
concentration of stickers and an unbound phase with a low sticker
concentration. Therefore, sufficiently strong barriers also lead
to lateral phase  separation and sticker aggregation.

In contrast to the entropic mechanisms mentioned above, the phase
separation  is caused by the line tension between bound and unbound
membrane patches. This line tension is induced by the potential barrier.
In order to understand the influence of membrane fluctuations, we have
to take into account that the transition value
$|U_{we}^*|$ of the contact energy increases with the temperature $T$
and decreases with the bending rigidity $\kappa$, see (\ref{Ucostar}).
As implied by (\ref{barcon}), membranes with more pronounced shape
fluctuations require larger potential barriers for a
discontinuous unbinding transition and lateral phase separation.
As in the first case studied in the previous section, membrane fluctuations
decrease the tendency of the membrane to phase separate.

\section{Conclusions}

In this article, we have studied one possible mechanism for
adhesion--induced phase separation of biomimetic membranes. The
mechanism is applicable to membranes which experience both
specific attractive and generic repulsive trans--interactions.
The specific interactions are taken to arise from sticker
molecules which are embedded in the membrane,  while the
generic repulsion may originate, {\it e.g.}, from electrostatic
forces between similarly and homogeneously charged membranes.
The effective trans--interaction obtained by an explicit summation of the
sticker degrees of freedom in the partition function is shown to exhibit
a potential barrier. This barrier induces a line tension between bound
and unbound membrane regions resulting in a coexistence of a sticker--rich
phase, characterized by  a small separation between membrane and
substrate, and a sticker--poor phase, characterized by a larger
membrane--substrate separation. Thermally excited shape fluctuations of
the membrane are shown to decrease the tendency for lateral phase
separation by reducing the potential barrier height.

Two different cases have been considered here. In the first case,
the generic trans-interaction between membrane and substrate is
assumed to have a minimum at a finite membrane separation and is
approximated by a harmonic potential centered at the minimum. The
specific trans--interaction of the stickers is taken to depend
linearly on membrane separation. This model   introduced in Ref.\
\cite{komuraAndelman} has been studied using a mean--field
approach. The adhesion was shown to shift the critical point of
the lateral phase transition. In the present article, the
concentration field $n$ of the stickers is described as a lattice
gas variable on a discretized elastic membrane. The phase
behavior is determined exactly in the absence of shape
fluctuations of the membrane. Fluctuations of the membrane are
subsequently taken into account by Monte Carlo simulations.

In the second case, the specific sticker trans--interaction is
modeled by a short--ranged square--well potential, and the generic
trans--interaction between membrane and substrate is assumed to be
a purely repulsive step function. This second case turns out to
be closely related to biomimetic membranes which contain stickers
{\em and} repellers, {\it i.e.}\ repulsive molecules which
protrude from the membrane surface \cite{lipo176}, since the
effective potential has the same functional form as in
eq.~(\ref{potbar}) and exhibits a potential barrier. As shown in
Refs.~\cite{lipo118,lipo124}, the membrane fluctuations can
`tunnel' through  a relatively small barrier but are trapped by a
relatively large one.

The mechanism for adhesion--induced phase separation as studied in
the present article is   distinct from several {\em entropic}
mechanisms which have been identified in previous works
\cite{weiklNetzLipowsky,lipo176}.  Examples of these other
mechanisms include   stickers with an increased lateral size
\cite{weiklNetzLipowsky},   stickers with an increased rigidity,
and stickers with attractive cis--interactions which are
renormalized by the shape fluctuations of the membrane
\cite{lipo176}. These entropic mechanisms depend strongly on the
rescaled potential range $(l_v/a)\sqrt{\kappa/T}$ of the stickers
where $l_v$ is the range of the square--well potential
(\ref{squarewell}), and $a$ is the size of the membrane patches.
Thus, for the entropic mechanisms, the  tendency for lateral
phase separation {\em increases}  with decreasing potential range
and/or     increasing temperature. This is in contrast to the
mechanism described in the present work which is governed by  the
potential barrier contained in the effective interaction potential
of the membrane.  In this  case, the shape fluctuations of the
membrane reduce  the potential barrier which implies that the
tendency for lateral phase separation  {\em decreases} with
increasing temperature.

Experimentally, the temperature--dependence of adhesion--induced
phase separation has not yet been studied. The presence of repulsive
lipopolymers in the biomimetic systems investigated in
\cite{albe97,klob99,gutt01,maie01} points towards a barrier--mechanism
for adhesion--induced phase separation as emphasized in this
article. Entropic
mechanisms, on the other hand,  might be relevant in the case of membranes
containing  oppositely charged lipids \cite{nard97,wong01} which induce a
tight
membrane coupling and small membrane separations below 4 nm
\cite{wong01}.

\subsection*{Acknowledgements:}
One of us (DA) would like to acknowledge partial support from the
U.S.--Israel Binational Foundation (B.S.F.) under grant No.\ 98--00429
and the Israel Science Foundation, Centers of Excellence Program.



\newpage


\newpage


\begin{thebibliography}{99}

\bibitem{liposack}
{\em Structure and dynamics of membranes: Generic and specific interactions},
Vol.~1B of {\em Handbook of
  biological physics}, edited by R. Lipowsky and E. Sackmann (Elsevier,
  Amsterdam, 1995).

\bibitem{helf78}
W.~Helfrich, Z.~Naturforsch. {\bf 33a}, 305  (1978).

\bibitem{lipo33}
R. Lipowsky and S. Leibler, Phys. Rev. Lett. {\bf 56},  2541  (1986).

\bibitem{bellEtAl}
G.~I. Bell, Science {\bf 200},  618  (1978);
G.~I. Bell, M. Dembo, and P. Bongrand, Biophys. J. {\bf 45},  1051  (1984);
G. I. Bell in
{\em {Physical basis of cell-cell adhesion}}, edited by P. Bongrand (CRC Press,
  Boca Raton, 1988), p. 227.

\bibitem{braunEtAl}
J. Braun, J.~R. Abney, and J.~C. Owicki, Nature {\bf 310},  316  (1984);
J. Braun, J.~R. Abney, and J.~C. Owicki, Biophys. J. {\bf 52},  427  (1987).

\bibitem{brui94}
R. Bruinsma, M. Goulian, and P. Pincus, Biophys. J. {\bf 67},  746  (1994).

\bibitem{albe94}
B. Alberts {\it et~al.}, {\em Molecular biology of the cell}, 3rd  ed.
  (Garland, New York, 1994).

\bibitem{lipo130}
R. Lipowsky, Phys. Rev. Lett. {\bf 77},  1652  (1996).

\bibitem{lipo176}
T.R.~Weikl and R.~Lipowsky, Phys.~Rev.~E {\bf 64}, 011903 (2001).

\bibitem{nard97}
J. Nardi, T. Feder, R. Bruinsma, and E. Sackmann, Europhys. Lett. {\bf 37},
  371  (1997).

\bibitem{wong01} A.\ P.\ Wong and J.\ T.\ Groves,
 J.\ Am.\ Chem.\ Soc.\ {\bf 123}, 12414 (2001).

\bibitem{albe97}
A. Albersd\"orfer, T. Feder, and E. Sackmann, Biophys. J. {\bf 73},  245
  (1997).

\bibitem{klob99}
A.~Kloboucek, A.~Behrisch, J.~Faix, and E.~Sackmann,
    Biophys.~J.~{\bf 77}, 2311 (1999).

\bibitem{gutt01} Z.\ Guttenberg, B.\ Lorz, E.\ Sackmann, and
A.\ Boulbitch, Europhys.\ Lett.\ {bf 54}, 826 (2001)

\bibitem{maie01} C.\ W.\ Maier, A.\ Behrisch, A.\ Kloboucek,
D.\ A.\ Simson, and R.\ Merkel, Eur.\ Phys.\ J.\ E {\bf 6}, 273 (2001)

\bibitem{lipo118}
R. Lipowsky, J.\ Phys.\ II France {\bf 4},  1755  (1994).

\bibitem{lipo124} A. Ammann and R. Lipowsky,
J.\ Phys.\ France {\bf 6}, 255 (1996).

\bibitem{weiklNetzLipowsky}
T.R.~Weikl, R.R.~Netz, and R.~Lipowsky,
 Phys.~Rev.~E {\bf 62}, R45 (2000).

\bibitem{komuraAndelman}
S.~Komura and D.~Andelman, Eur.~Phys.~J.~E, {\bf 3}, 259 (2000).

\bibitem{brui00}
R. Bruinsma, A. Behrisch, and E. Sackmann, Phys. Rev. E {\bf 61},  4253
  (2000).

\bibitem{lipo154}
R. Goetz, G. Gompper, and R. Lipowsky, Phys. Rev. Lett. {\bf 82},  221  (1999).

\bibitem{canham} P.B.\ Canham, J.\ Theor.\ Biol.\ {\bf 26}, 61 (1970).

\bibitem{helf73}
W.~Helfrich, Z.~Naturforsch.~{\bf 28c}, 693 (1973).

\bibitem{rlEPL88}
R.~Lipowsky, Europhys.~Lett.~{\bf 7}, 255 (1988).

\bibitem{binder} K.~Binder and D.W.~Heermann,
{\it Monte Carlo simulations in statistical physics} (Springer, Berlin, 1992).


\end{thebibliography}
\end{document}